\documentclass{nature}
\usepackage{epsfig,verbatim}
\usepackage{graphicx}
\usepackage{amsmath}
\usepackage{revsymb}
\usepackage{amssymb}

\usepackage[caption=false]{subfig}

\allowdisplaybreaks

\mathchardef\ordinarycolon\mathcode`\:
\mathcode`\:=\string"8000
\def\vcentcolon{\mathrel{\mathop\ordinarycolon}}
\begingroup \catcode`\:=\active
  \lowercase{\endgroup
  \let :\vcentcolon
  }
\newcommand{\ket}[1]{|#1\rangle}

\title{Pretending to factor large numbers on a quantum computer}

\author{John A. Smolin$^1$, Graeme Smith$^1$ and Alex Vargo$^1$}

\begin{document}
\maketitle
\begin{affiliations}
\item IBM T.J. Watson Research Center, Yorktown Heights, NY 10598, USA
\end{affiliations}

\begin{abstract}

Shor's algorithm for factoring in polynomial time on a quantum
computer\cite{Shor} gives an enormous advantage over all known
classical factoring algorithm.  We demonstrate how to factor products
of large prime numbers using a compiled version of Shor's quantum
factoring algorithm.  Our technique can factor all products of $p,q$
such that $p,q$ are unequal primes greater than two, runs in constant
time, and requires only two coherent qubits.  This illustrates that the
correct measure of difficulty when implementing Shor's algorithm is not
the size of number factored, but the length of the period found.
\end{abstract}

\section{Introduction}

Building a quantum computer capable of factoring larger numbers than
any classical computer can hope to is one of the grand challenges of
computing in the 21st century.  While still far off, there have
already been several small-scale demonstrations of Shor's
algorithm\cite{Chuang2001,Lanyon2007,Lu2007,Politi2009,Factoring21,Lucero2012}.
Someday soon a quantum computer may factor a number hitherto
unthinkably large.  Such a device would most likely have to be a fully
scalable fault-tolerant quantum machine, capable of carrying out any
task a quantum computer could be asked to do.  Thus, a large
factorization would be convincing proof that one has built a practical
quantum computer.  Until such a time, more modest goals must suffice. 
The experiments mentioned above have factored numbers no larger than 21.
Here we will show how current technology can demonstrate significantly larger
factorizations.

We begin with a review of Shor's algorithm.  Given an integer $N=pq$ with
$p,q$ distinct primes, one proceeds as follows:
\begin{enumerate}
\item Choose (at random) an integer $ 0< a < N$.
\item Compute the greatest common divisor (GCD) of $a$ and $N$. This can
be found efficiently using the Euclidian algorithm\cite{Euclid}. If it
  is not 1, then ${\rm GCD}(a,N)$ is a nontrivial factor of $N$.
  Otherwise go on to the next step.
\item Choose $S\equiv 2^s$ such that $N^2 \le S < 2 N^2$.
Construct the quantum state
\begin{equation}
S^{-1/2}\sum_{x=0}^{S-1}\ket{x}\ket{0}
\end{equation}  on two quantum registers, the first is $s$-qubits and the second is
$\log N$ qubits.
Note that in the literature $x$ and $a$ sometimes have their meanings
interchanged.

\item Perform a quantum computation on this state which maps $\ket{x}\ket{0}$
to $\ket{x}\ket{a^x \mod N}$.  This is the slowest step, but 
can be done in time $O((\log N)^3)$.
\item Do the quantum Fourier transform on the first register, resulting in the
state
\begin{equation}
S^{-1} \sum_x \sum_y e^{(2\pi i /S)xy}\ket{y}\ket{a^x \mod N}\ .
\end{equation}
This step requires $O((\log N)^2)$ time, which is much less than the
modular exponentiation of the previous step.

\item Measure the first register to obtain classical result $y$.  With
  reasonable probability, the continued fraction approximation of
  $y/S$ or some $y'/S$ for some $y'$ near $y$ will be an integer
  multiple of the period $r$ of the function $a^x \mod N$.  The GCD
  algorithm can then efficiently find $r$.

\item If $r$ is odd, or if $a^{r/2} = -1 \mod N$, go back to step 1.  Otherwise,
${\rm GCD}(a^{r/2}\pm 1,N)$ is $p$ or $q$.

\end{enumerate}

\begin{figure}
\centering{\includegraphics[height=1.3in]{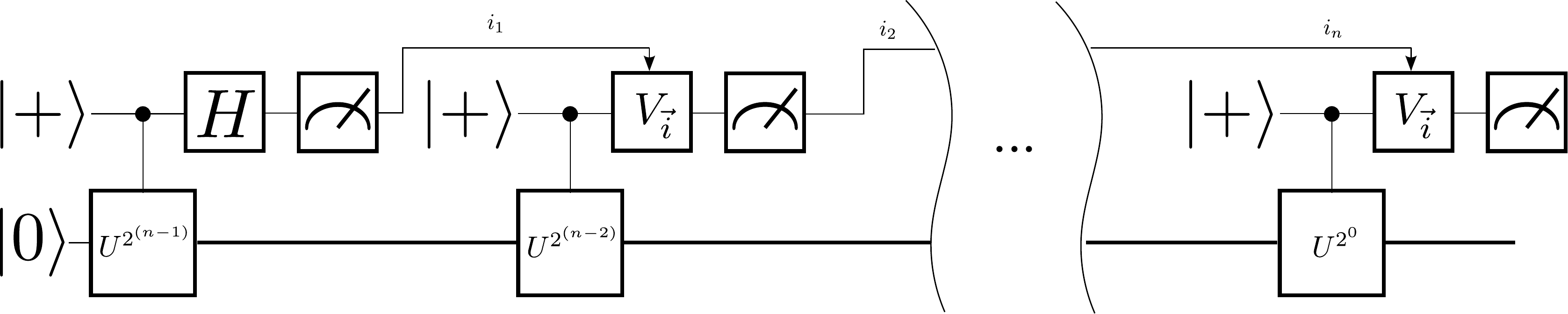}}
\caption{Circuit for Shor's algorithm using the semi-classical quantum
  Fourier transform.  At each stage a $\ket{+}$ state is prepared.  It
  is used as the control input on a controlled unitary $U^{2^{n-1}}$ for the $n$th
bit of the readout, with \mbox{$U\ket{y} =
  \ket{a y \mod N}$}.  Next, the gate $V_{\vec{i}}={1\ \ \ 0
    \choose 0\ e^{i \phi}} H$ is applied and then the qubit is
  measured.  $H=\frac{1}{2}{1\ \ 1\choose 1\ \!-1}$ is the Hadamard
  gate and the phase $\phi$ is computed as a function of all previous
  measurement results (see Ref.\cite{GriffithsNiu1996}).  The first
  time there is no phase so the Hadamard is used.  The process is
  repeated $n$ times to read out $n$ bits of precision of the Fourier
  transform.}
\label{fig:recycling}
\end{figure}

Significant optimization of the basic algorithm has been achieved.  As
described, roughly $3 \log N$ qubits are needed.  In fact, this can be
reduced down to exactly $2 + 3/2 \log N$ qubits\cite{Zalka2006}.  A
significant part of the reduction is to replace the first ``$x$''
register with a single qubit.  This was shown to be
possible\cite{MoscaEkert1999,ParkerPlenio2000} and uses the fact that the bits of the
quantum Fourier transform can be read out one at a
time\cite{GriffithsNiu1996}.  The use of this semi-classical Fourier transform
has become known as {\em qubit recycling}.  A circuit using qubit recycling is shown
in Fig. \ref{fig:recycling}.

\section{Compiled Shor's Algorithm}

All experimental realizations of Shor's algorithm to date have relied
on a further optimization, that of ``compiling'' the algorithm.  This
means employing the observation that different bases $a$ in the
modular exponentiation lead to different periods of the function $a^x
\mod N$.  Some of the periods are both short and lead to a
factorization of the composite $pq$.

In 2001, the composite 15 was factored\cite{Chuang2001} using two
different bases, an ``easy'' base ($a=11$, resulting in a period of
2), and a ``difficult'' base ($a=7$, with a period $r=4$). Neither is
fully general and this allowed the factorization to take place on a
seven bit quantum computer, when the best known uncompiled algorithm
would require 8 bits ($2 + 3/2 \log N$ bits as per
Zalka\cite{Zalka2006}).  Other factorizations of 15 have since been performed using
other architectures\cite{Lanyon2007,Lu2007,Politi2009,Lucero2012}.
More recently, 21 has been
factored\cite{Factoring21} using just one qubit and one qutrit (a
three-level system).  In this case $a=4$ is used, resulting in a period 
$r=3$\footnote{Note that Shor's algorithm normally fails when $r$ is odd since
$a^{r/2}$ is irrational in general.  Here, since $a=4$ is a perfect square, 
this problem does not arise.}.  These results are summarized in Table 1.

\begin{figure}
\centering{\includegraphics[height=2in]{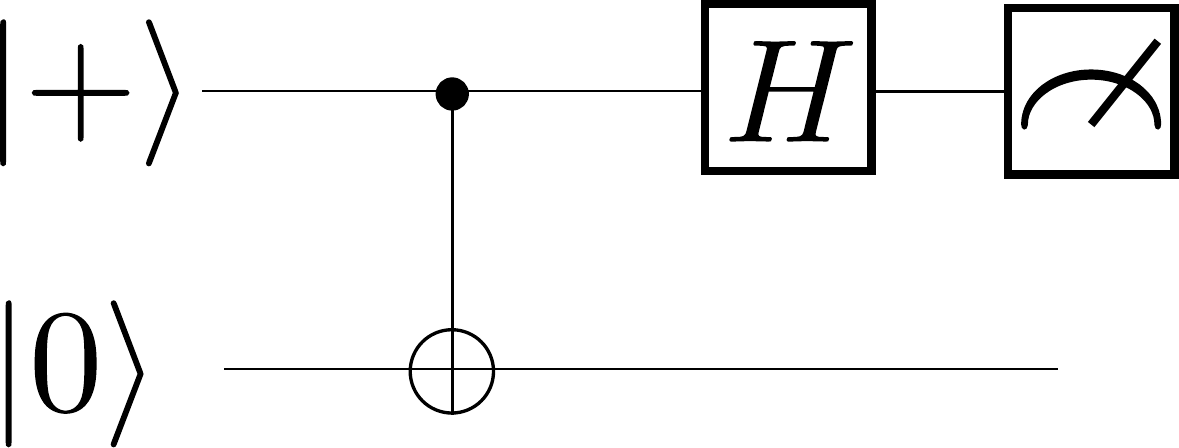}}
\caption{The circuit for the fully-compiled Shor's algorithm.  The modular 
exponentiation is the single controlled-NOT, and the quantum Fourier transform
is a Hadamard gate.  }
\label{fig:compiled}
\end{figure}

Recently, Zhou and Geller showed\cite{Factoring51and85} how to find
$a$'s with small periods for products of Fermat
Primes\cite{FermatPrimes} (primes of the form $2^{2^k}+1$).  Here, we
go substantially beyond this idea and show that \emph{any} composite
number $pq$ has compiled versions of Shor's algorithm that can be run
on a very small quantum computer.  In particular, we show that there
always exists a base $a$ such that $r=2$.  Then, the second register
need only hold two distinct states and the computation can be
performed using only two qubits.  In this case, the $U$ needed in the
circuit from Fig. \ref{fig:recycling} reduces to a controlled-NOT
gate.  Furthermore, only one stage of the circuit is required since
all powers of $U^{2^n}$ are the identity except for $n=0$.  The
compiled circuit is shown in Fig. \ref{fig:compiled}.


In order for the second register to need to hold only two distinct states,
we must find a base $a$ such that $a^2=1\mod pq$.  The Chinese remainder
theorem\cite{ChineseRemainder} tells us that
\begin{equation}
\label{CR}
a^2 =1\!\!\mod pq\ \Leftrightarrow\ a^2=1\!\!\mod p\ \ \mathrm{and}\ \ a^2=1\!\!\mod q
\end{equation}
for $p,q$ relatively prime.
By construction 
\begin{equation}
a\equiv \pm p p_q \pm qq_p\ \textrm{has}\ a^2=1\!\!\mod p\ \textrm{and}\ a^2=1\!\!\mod q 
\label{CR2}
\end{equation}
where $p_q$ is the multiplicative inverse of $p$, $\mod q$ and $q_p$ is 
the inverse of $q$, $\mod p$.  Then (\ref{CR}) tells us $a^2=1 \mod pq$.
These inverses can be found efficiently using the extended Euclidian algorithm.
There are 4 solutions of (\ref{CR2}) corresponding to the signs.  Two of these
will be trivial, $\pm 1$ and the other two will be bases resulting in
compiled Shor factorizations with a period of the function $a^x\!\!\mod N$ 
having period 2.

\begin{table}
\centering{\begin{tabular}{cccc}
\hline\\
$N$&Qubits needed (Zalka\cite{Zalka2006})&Qubits implemented&Qubits compiled\\
\hline\\
15&8&7 [Ref.\cite{Chuang2001}],4 [Refs.\cite{Lanyon2007,Lu2007}],5 [Ref.\cite{Politi2009}],3 [Ref.\cite{Lucero2012}]&2\\
21&10&$1+\log 3$ [Ref.\cite{Factoring21}]&2\\
RSA-768&1154&2$^\dag$&2\\
N-20000&30002&2$^\dag$&2\\
\hline\\
\end{tabular}}
\caption{Number of qubits required for Shor's algorithm and
  experimental results.  RSA-768 and N-20000 are available in the
  supplementary material.  $^\dag$Quantum version to be completed.  Classical
  version with one random bit has been performed, see Section
  \protect\ref{experiment}. }
\end{table}

\section{Experiment}
\label{experiment}

\begin{figure}
\centering{\includegraphics[height=1.8in]{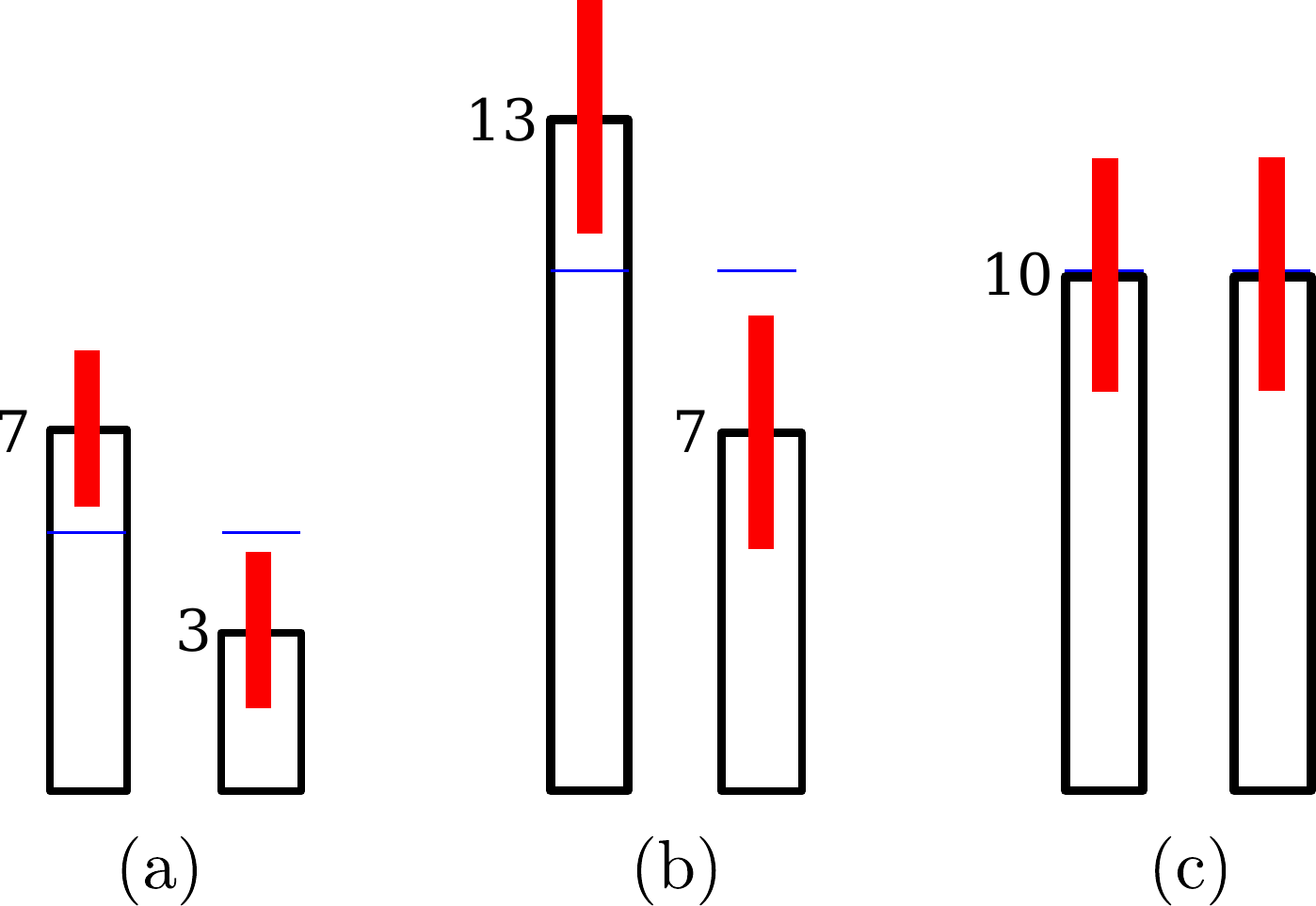}}
\caption{Experimental data from unbiased coins.  (a) A 1998 US quarter 
was tossed 10 times to factor 15.  (b) A 1968 US penny was tossed 20 times
in order to factor RSA-768.  (c) A 2008 US Oklahoma commemorative quarter
was tossed 20 times to factor N-20000.  One-$\sigma$ error bars are shown.}
\label{fig:data}
\end{figure}

A future version of this preprint will include experimental data using
two superconducting transmon\cite{Transmon} qubits.  In the meantime,
we perform a simpler experiment.  We employ a further optimization not
used in previous experiments.  Observe that in the circuit in Figure
\ref{fig:compiled}, the second qubit is never measured.  In fact, what
is created by the controlled-NOT is a maximally-entangled state, half
of which is simply discarded.  The resulting state of the first qubit
is therefore maximally mixed.  Due to the unitary equivalence of
purifications, if we create a maximally mixed state in any way at all,
it is entangled with some system in the environment.  A maximally
mixed state is unaffected by the Hadamard gate, so this too is
unnecessary.  We can therefore produce the appropriate probability
distribution at the output by tossing an unbiased coin.
Fig. \ref{fig:data} shows the data for factoring 15, RSA-768, and
N-20000 using this method.

\section{Conclusions}
Of course this should not be considered a serious demonstration of
Shor's algorithm.  It does, however, illustrate the danger in
``compiled'' demonstrations of Shor's algorithm.  To varying degrees,
\emph{all} previous factorization experiments have benefited from this
artifice.  While there is no objection to having a classical compiler help
design a quantum circuit (indeed, probably all quantum computers will
function in this way), it is \emph{not} legitimate for a compiler to
know the answer to the problem being solved.  To even call such a
procedure compilation is an abuse of language.  

As the cases of RSA-768 and N-20000 suggest, very large numbers can be
trivially factored if we were to allow this.  For this reason we
stress that a factorization experiment should be judged not by the
size of the number factored but by the size of the period found.
Current experiments ought to be viewed instead as technology
demonstrations, showing that we can manipulate small numbers of qubits.
In Ref.\cite{Factoring21}, for instance, it was shown that
intentionally added decoherence reduced the contrast in the data, 
a hallmark of a quantum-coherent process.  All the referenced experiments
are important tiny steps in the direction of building a quantum computer,
but actually running algorithms on such tiny experiments is a somewhat
frivolous endeavor.

\section{Acknowledgements}
We acknowledge support from IARPA under contract no. W911NF-10-1-0324 and
from the DARPA QUEST program under contract no. HR0011-09-C-0047.
All statements of fact, opinion or conclusions contained herein are
those of the authors and should not be construed as representing the
official views or policies of the U.S. Government.

\bibliographystyle{unsrt}

\pagebreak
\bibliography{pretend}

\section{Supplementary material}
We have factored RSA-768:
\begin{align}
\nonumber\mathrm{RSA-768}=&1230186684530117755130494958384962720772853569595334792197322&\\
\nonumber &4521517264005072636575187452021997864693899564749427740638459&\\
\nonumber &2519255732630345373154826850791702612214291346167042921431160&\\
\nonumber &2221240479274737794080665351419597459856902143413&\\
\nonumber=&3347807169895689878604416984821269081770479498371376856891243&\\
\nonumber &388982883793878002287614711652531743087737814467999489&\\
\nonumber       &\times 3674604366679959042824463379962795263227915816434308764267&\\
\nonumber       &6032283815739666511279233373417143396810270092798736308917&
\end{align}
which can be done using
\begin{align}
\nonumber a=&1029031793302493258003488818376905875264575120178567995715921117383374&\\
\nonumber &0637809554762657146559655560974877155097084531342124720712415517107376&\\
\nonumber &6764612501767199553731974973903504534358652759946682893508255761840004&\\
\nonumber &7627481255809299529939\ \ \mathrm{or}&\\
\nonumber a=&2011548912276244971270061400080568455082784494167667964814013347683523&\\
\nonumber &3672630818125303054623423037190224096523432092963345312131577459271606&\\
\nonumber &3902143490245030584823163722635383870725074621773650339655236462265303&\\
\nonumber &792116204047602613474.&
\end{align}

We have also factored N-20000:
\begin{align}
\nonumber\mathrm{N-20000}=&3545872995518995320162216211194750088254004994975476505069782010092562&\\\nonumber &
6836455218077570931827645265231608998412594091501169267869932212762643&\\\nonumber &
7952340916754401101694942393398796986114303891110264606442388558651298&\\\nonumber &
8969181675291250725852835615290690458331490854580449715364904882361230&\\\nonumber &
3234271985470197470205533036194402979614663510505596285870603463725841&\\\nonumber &
3294436223612733881541244957256787578366712932408954804647829260155941&\\\nonumber &
1444057806766879951233726684092407204779854854431536554954640444167496&\\\nonumber &
3844339791512553551092864281417298489668114677339033398078956576863772&\\\nonumber &
6354676678753024806722863674385802601319484610034919950211309093009540&\\\nonumber &
5949579561782282227038687651707862691988512957531427525039477044137682&\\\nonumber &
2111920721522148087696432046839747743667973894759586729469351498803425&\\\nonumber &
8962044592322773864737596716914996302081975587285523928985393001093345&\\\nonumber &
3532923398023447380352816430049211499186658152360647169892936850272053&\\\nonumber &
0977920879248353643585686181315512898374697516865004762004956941084908&\\\nonumber &
8691670020716296801864932328763094256328083479904065324168359574206973&\\\nonumber &
2449257885376492437902211127107636812954877711352430874473152262662092&\\\nonumber &
1393167409565204821691194551281949537731218793132259214278948500148983&\\\nonumber &
3344871961435511910047360734736812014832840402983502352620746227654188&\\\nonumber &
4821175368994484765461270105712673387687224075345896614234794751348388&\\\nonumber &
4010462683805164134781712394881679848878483404785832664176081990699296&\\\nonumber &
6948006619580318497838879486674968615405796077696761351965322716784614&\\\nonumber &
7871304649614628515620902547289543420206114653340656249551468555247381&\\\nonumber &
8240712032875920730690848056307144691354557092267705269040098564622603&\\\nonumber &
2392695881701195013830935107656843989911154471193950278950107730349221&\\\nonumber &
4291514924457659248459718543654416911772236786502398286527705256699729&\\\nonumber &
7748537137527061667330019454152656165205315539492875423667810539723325&\\\nonumber &
7022274282342200952354298403423449776728858821486783470468226452695002&\\\nonumber &
9965935276141786952752454295813011211063314171867861464423084956383254&\\\nonumber &
9250134337994353366719174074821153482037985373069187708394252853900085&\\\nonumber &
4621063052758727903887688129106643403113169705619311899904858481305060&\\\nonumber &
6794594308020321189052847966275370327401680437045200772702267192976527&\\\nonumber &
8022092616552738346118012463149731620383592103183116018168719777412966&\\\nonumber &
4680141180854170139904809370511462062235669898350272862689886145264779&\\\nonumber &
0432452754012715374184451464019524311183243733549609931901183953471741&\\\nonumber &
9682604800709871547108871240628342851823772296896287863147378366801691&\\\nonumber &
4146092518979691440024256035159221892354831540906789710253791305057945&\\\nonumber &
2513618763200501337192664729579391825335186059463895910648093389943606&\\\nonumber &
9023498084198924775533135861627217005509336538759667215543315715133914&\\\nonumber &
6545373118930562292794711875572014653533038100140905439252036847724360&\\\nonumber &
4084047742325464019428788002314216269103442810812782163053880781848891&\\\nonumber &
0976756747825489367195795173775440332520928104948661596841414261882855&\\\nonumber &
0508084652551183976898343978994771023762438504588172953877252482623051&\\\nonumber &
5919224994427013165388804662115091908342462138499805129411607905132572&\\\nonumber &
4944127375531074567010781507626054608117591697601660015955369691615690&\\\nonumber &
7299677384025069458631358698179643840467067437522628304697721524957084&\\\nonumber &
5459564669588865188810573355621573095176830937715876692732697106901427&\\\nonumber &
4380431733357921633228621406293959828148048345567080697512994682036864&\\\nonumber &
2273953787724284705267144103683177203987376883047129500494898696824232&\\\nonumber &
9532558731137573475789782578092773219041672171778789766725155223594083&\\\nonumber &
7800940240714338777859756549429556992790225005076766692433272372515832&\\\nonumber &
5468083394216073893463198581373135761145571139995993471008944898117555&\\\nonumber &
7915110867759507303536690925541985870318508355007476077608513903304792&\\\nonumber &
3425828007998127685805726212734915981827589500867306063576535238956394&\\\nonumber &
6793395313450971874893873569704161228657978315778567085288463502009723&\\\nonumber &
9226270256866094212840482256101905075703352067871310155247660290496318&\\\nonumber &
4692638571607375563996900070807339193161786801869496736152219039495203&\\\nonumber &
5125269783067194766221272323906814360475290868630208923012179164036085&\\\nonumber &
9837356438242466155230648491979782175513982312032742419908008157043521&\\\nonumber &
0580254128091700530671326828205445719642746517340874658991929885960264&\\\nonumber &
7517647709753226504353677801754694678666697336302625047413771645854127&\\\nonumber &
0547527549069602306014337414254789438208964165690841775476825804165152&\\\nonumber &
5304663247060964001143198123077855794438185702605450833342983353775818&\\\nonumber &
5889572134166889296081487694243723204621491394952553741414901724430252&\\\nonumber &
8855616909378500458286429434922098019136721665788293210513458527694013&\\\nonumber &
2216643750671547230453049952426398499419053393702254349492760414472993&\\\nonumber &
0642315675678036189230393221614960576525703335669945041027365152055579&\\\nonumber &
5744126705024976289705383585305834157781770550223289305932646500163282&\\\nonumber &
0116464325733138774212045809506632783143002474700159635962123125236917&\\\nonumber &
4409101844438736437858315266946108779141151800556377477506707469599007&\\\nonumber &
3102992317593007543603235409748865074021408056842928974647832116695824&\\\nonumber &
5583057892303965553619740076982226149049986747523605054240483216161269&\\\nonumber &
5845735514152275026785133897200124109426854987571162068144453489055014&\\\nonumber &
5864970551342995682354407222447173109382329057323719589044948028095133&\\\nonumber &
4711147625611503091127337452960350812003389806489437389335138585299176&\\\nonumber &
8503183902737153079057664904077587776062549909997744891283441109343053&\\\nonumber &
9710212166551627117192070283735855586679692418134287944889037963176608&\\\nonumber &
8271018129029021585609464256357290486028253978413567783992747663703346&\\\nonumber &
4998930738742796686640128620923272630180172545665881974962283565208378&\\\nonumber &
8146895814608717113146637672212361020062046644744553178935511572078813&\\\nonumber &
4289685668894167388898155080721001862530295759444117033443834883115100&\\\nonumber &
0650673799192981407882986789587653580546902745162832189506923934649727&\\\nonumber &
3468283522712516205990343243813973588848694703996512441662433168667879&\\\nonumber &
1132220512140731023345171452862842540349609656231516812698183384579682&\\\nonumber &
7082647120785562227006525356718270612869805415804935095162605716061024&\\\nonumber &
4348906073326891005641611113422562155003437331976928597724313112437132&\\\nonumber &
9726226175793616722507421635694278871177878800928003913785632249721075\\\nonumber &
9&\\
\nonumber=&
1920007954675520804378850214567096264456508253959234467727111623305767&\\\nonumber &
7374265185330406795304487914263190662072636603750445244958319471906305&\\\nonumber &
0546199924938408954543563150180797425730245513649150981962229175055926&\\\nonumber &
5639970111115853501252719600314314117110756537592110376119248953599128&\\\nonumber &
3899768868030610997197057941193788301384925356840915526630432721419360&\\\nonumber &
4879721566536637493968167315831870622227408193652848736544484648059007&\\\nonumber &
8879203308810158844503374309748890063940900027992536261993320143970650&\\\nonumber &
4546707853810288254711599606525009903719426231146990556873785020501023&\\\nonumber &
6921309632407499531263607144034119343854298380373180935294905206027753&\\\nonumber &
3316164523654367390074206381449205517815424087196134280305359588385106&\\\nonumber &
0148885682715175492354454895919496808226041535676588909998629212717595&\\\nonumber &
7252045335576196896285685325032772197585932310039602104286674497751751&\\\nonumber &
3735511346429161240577093834733605413619341789239569628496964487858741&\\\nonumber &
9397025118511173835678856918497340476843981151822753936393438415284562&\\\nonumber &
2595327259462749546191534394708028267948168734606287494477015179631963&\\\nonumber &
3128988988026887797548866985684538406348675891721165893482443941143046&\\\nonumber &
9198897765504664490525671701566860302600604994891420505724328362982218&\\\nonumber &
5869500891998443885693582489052192339402286011515307511936291172078611&\\\nonumber &
5369435318064427753875206862604881427693151447822040695630536054475353&\\\nonumber &
3757533602651731156682480712806900803919574751471016869541117827746736&\\\nonumber &
5536021391516685181241161433621755613145667340298230128682616323239082&\\\nonumber &
3357619809691615460088802326462712010390810975054420845793817460736705&\\\nonumber &
2466388857583094796671645765497531297255387079830529056697297735365569&\\\nonumber &
8272054576608357098630860329108094169459004787030168446544659938078598&\\\nonumber &
5561989769518993444500177834799824245748273057983371486640604699710487&\\\nonumber &
4770533916896193700019269730005874534486128087333932120639645595377582&\\\nonumber &
1046929424126942165409602171262376117287841141661522578005357658615376&\\\nonumber &
9418694236300911166268868347387404024568881764404915174437824364735237&\\\nonumber &
0193628308367963132731910962182993571571649918139994774563087499048797&\\\nonumber &
4894318231809141370840393798152696022983479657048672802741548348653208&\\\nonumber &
7256329064122941199182695807446254219961505997352333010564207142241119&\\\nonumber &
8854298777617828048368908767864646729247705615907317577720222897361292&\\\nonumber &
2113013794638075033833008149473852130731813326482758226200609938833244&\\\nonumber &
6141896781756368023768571213708719369689067047153100989591494973353316&\\\nonumber &
2552717367854676463006032590457099891058545676094521480005970410137677&\\\nonumber &
7440439008115126683654630139325103582587972621504249537137741947041111&\\\nonumber &
9615032757065716791252275469697876322879301632967912158645043740191488&\\\nonumber &
2867351034199861481812601249004704113433829496472388970117012344391785&\\\nonumber &
5214895963990645885365973014138246198888626501569449226520526513439295&\\\nonumber &
8062536359277041944045049865774322631065584998563338496575099527026242&\\\nonumber &
6259463783610590851095864429551274069878588826066040975266553379144063&\\\nonumber &
8233758599521003191447351046978611683337066197676311673983746499230763&\\\nonumber &
4653512156501078415246903395966620172135755433690037220483180632495240&\\\nonumber &
3&\\
\nonumber&\times 
1846801200424320058688545075138064936637978107808827998771506577923856&\\\nonumber &
3512240821642698240746519848908888538863314146811777562735349842383990&\\\nonumber &
5670861632022702344791536050984503349712490164967553980037733570521519&\\\nonumber &
7292378714384123838303682671000232696966397317822765773532584800195449&\\\nonumber &
5368601567970009628561986870825780394439849598274041251373721318688499&\\\nonumber &
1453459939131218282383837906067367719975956305299253544862766250674360&\\\nonumber &
8998556300881542429404051609300094861204547432754918123186836075613988&\\\nonumber &
2688116428895169498745291929770561287646470297265115687387483895368013&\\\nonumber &
6199972574717643411232857324289391893113632565488962345764030022326064&\\\nonumber &
3637140144476343583887275463815046470469103711530263500124285402956246&\\\nonumber &
3762807770858497146737449843628571647621934352570516005139021147129387&\\\nonumber &
7254093410928119195530593476356748117307256152959591420794993258199178&\\\nonumber &
1288658087938767149531957135085795481815496115596710030791084073641958&\\\nonumber &
2481531196768438309249368833213399954953500171444271211952594455954004&\\\nonumber &
4777528236825191633941191910648017375990757253500191633238179812152213&\\\nonumber &
4810147069844617788669286684585073469838164562197551333229843987638772&\\\nonumber &
5945072519444544955988951672961243045362454176002958543468169547377785&\\\nonumber &
2118090983826385814893432815255415166551466218444312659395947517896476&\\\nonumber &
5227972655432667418542843772274135952480031558350771481551530387576682&\\\nonumber &
1041672499278142585801212148975733472720705737412541961935461361686616&\\\nonumber &
3508594760113802885442359428129640983557807310598864326698242112052172&\\\nonumber &
7692308141139830587045985601413400461409792071967578041001169056838821&\\\nonumber &
3000815223385679073522081944979933765331520218769929503537517594373554&\\\nonumber &
2191945665321791135058780449288146905346066385682543573570427539134224&\\\nonumber &
8574920595472902508707932903893351324379104610323193146701342949494634&\\\nonumber &
5040389611215775286692930984369581189082963880754080550863065745759611&\\\nonumber &
0034605626644345715399005851440737729661919797615755756108761930145637&\\\nonumber &
1933646822037580035275319476579099861322932643009662086631439248739876&\\\nonumber &
3971070581855559310526540801023876149579880675587365810942898804964034&\\\nonumber &
3641324793480945404915907217691465923415922903253537229239282128442771&\\\nonumber &
0130154358888273026907442099494780142517980536411086833639103946434790&\\\nonumber &
8079529206806056747235446064339813017857678116498351789018935064726714&\\\nonumber &
1205610121845710425453033644840293119085758895565396129325889160610871&\\\nonumber &
0001861154150674228966256352963445705488645252971334830788092158736981&\\\nonumber &
8548442060025794105215148425586279208114977130525142260634945369806224&\\\nonumber &
2951205784856742549955754122223419902899444511949432902275576616515852&\\\nonumber &
0298429438590678595614130944793904427593963617838940488779864894652743&\\\nonumber &
8929581534223224707905264709473535937313287694876857699278157956559231&\\\nonumber &
8143880758180364640966492445434588823620859839153173651135758856091438&\\\nonumber &
9042211094006903936734184536041517653820628447454164726119878611731520&\\\nonumber &
7560909778253371974487292012272567041087110701294986228504764820849761&\\\nonumber &
8573495632009171816411003696358627589713769962223373373249199890293069&\\\nonumber &
3381292825531337580852005811880055274693356863879867959600226865133785&\\\nonumber &
3&
\end{align}
using
\begin{align}
\nonumber a=&7768655490606911685159539689674850887282976739122743910922331484127768&\\\nonumber &
1101598247891527093924276297376331925480726454586027525740679667172791&\\\nonumber &
7217555058225806037812849768665258620363753772039717327412351869425858&\\\nonumber &
1904615133725741922663209052524169553197120804112291503080775196150179&\\\nonumber &
7481370912095756084237524803358135914101301946505352067303719701391298&\\\nonumber &
9633573034906117288024060883067970149350028279260607246975247889745922&\\\nonumber &
3602402225861505003313572304306905447606268313967386457264305604064558&\\\nonumber &
3367157507463118189262419467218612739212141448472566228406670407624687&\\\nonumber &
8486221019468246096874693256497011232185912639187623568434411158159227&\\\nonumber &
9425466290979790406124616963745434377993553169254764308587519584763283&\\\nonumber &
9908026660383906748955595904366133542000684895868104840199006615604508&\\\nonumber &
6894705435816406339881113921112298180391479715714729193061759403575720&\\\nonumber &
8867081097073744756977406957609445562091547506991676140472056012953296&\\\nonumber &
2991197632461565608050410540721478822950371795128621835949856963702166&\\\nonumber &
6224371370566744727252209694300253221185119722276569918131887004160566&\\\nonumber &
6511787743518341447546145449288405082375973635363541085217058550863051&\\\nonumber &
8839480229319048989879273899633182443157305887861149360197255794463171&\\\nonumber &
6709795940333192144664906516857203200395107084095369898155413386678738&\\\nonumber &
6681168343198907068353082927344962384696724403480382570588845441306312&\\\nonumber &
4877488397533501762174021757894540809497389965049570340694529853685363&\\\nonumber &
5635561992462676301235044711032542520182923630452963480775815035924535&\\\nonumber &
7407585970491234849229578971342439324657154333296083033927205778629383&\\\nonumber &
5933456610546742384149125391550133804528633362962890384572999592010518&\\\nonumber &
0577144064765208873501988645792223632904062971555444609318425080214212&\\\nonumber &
1184384417379276316498802813310085391510387559738744828556432487157726&\\\nonumber &
3536507192121026667516804010636699140175025481989452339630498257434762&\\\nonumber &
2011645970891564118158388911018971494376698527379018682192193839963988&\\\nonumber &
7870288971199058974835805676911081631660460528900275797960813065304569&\\\nonumber &
0640712653000612283793619074450258251735993639549632201530232832672693&\\\nonumber &
0306072350183855571253723653224706512755551461545529530725863741504341&\\\nonumber &
3397800741167746930322518549547430637708304127596783059385105277735737&\\\nonumber &
8482614301436370566936056058292203927221644388969239418286238968110267&\\\nonumber &
6882022887031668005695897091456731281290012865102080804589769739283386&\\\nonumber &
6600783860966017449648782410435397842634029430241490523392488159825914&\\\nonumber &
2672988420057289152775271944037905438583915655458922682125310128010837&\\\nonumber &
8858195260462227622175371396037284614133691864051700456905005940560843&\\\nonumber &
7249620502937147500597335949237875865659192847456273342312199826615359&\\\nonumber &
3897479465751856486775507185255887953484663756182652027623502598249541&\\\nonumber &
7409272930036883372543850327216737105418081476434270190610860338784663&\\\nonumber &
8831793483858477942469102133722623755960073496030454924036170912597889&\\\nonumber &
9182900979031855786297551515594299636397234876879796406211203590974956&\\\nonumber &
5597450433913377786561630300563507193714756659799573805984190638740668&\\\nonumber &
5778742626218992622016789035980975517276693684384338525402268079025529&\\\nonumber &
8415114337585790907673062652756385030374552404077114473245680271217408&\\\nonumber &
9679493206197672107500097731335581124262543288672783558407576576492367&\\\nonumber &
6061266251606506380783504941140293730801242198904583437784827122262226&\\\nonumber &
0304661756503482059134545468144302111738047206642154563479783777942934&\\\nonumber &
3287982083407105130822823601636668947774417556085011322360653035736163&\\\nonumber &
8498115198691052587526986800586478769573221961428294708523960609978921&\\\nonumber &
7210063876356625581417040325129360350943722331331576915147806841402642&\\\nonumber &
3062158428777261815692700687258586284011890534327010727310072426147618&\\\nonumber &
0032876449552439635119789614284359266300337555774891673793070973247162&\\\nonumber &
0687437474665913482817922512150641901227130850761227264795192649287354&\\\nonumber &
5545706011153733353219932601126600744346385464543947722265811810437706&\\\nonumber &
0305146715037086881369296321271327609689538672943344965051322218714216&\\\nonumber &
1394232745606515270931687434902610371736891381305010709357121801240219&\\\nonumber &
3451083561414756511499854549570627949301381511627600160894856144791642&\\\nonumber &
2748827283445788727633245920886635555532002527442395382483953810070922&\\\nonumber &
5900716044726465601327434337177980671149536034388615492302914701460003&\\\nonumber &
9026949235041267254916713419859156893361211053514503998384178397174028&\\\nonumber &
8683858210654145665577414609651511299226334367068466393454946132367415&\\\nonumber &
4688725976109664081661553595940519168967156394456608058049105657335771&\\\nonumber &
5448393194647003519362130711792172914134009262829896596618545955389312&\\\nonumber &
4015649980327075605354431029643598811641055399308773452799220589454272&\\\nonumber &
0952174425508133189754881919156955811929193191966781308816208523443966&\\\nonumber &
8891300902123846442621655553813255010016293418231264858438746981364764&\\\nonumber &
3536090363721369244372494220169902044604814634046636718112295834871166&\\\nonumber &
1412468088215513329401649126838868492020943512718719167525882772631465&\\\nonumber &
8162771799678000818183545474093699428524225986161319723996003357752179&\\\nonumber &
4715347799726928642560407209901126345877182094113950322701042222509684&\\\nonumber &
0885336799345183903647399240608181342871848835612894730505009535329303&\\\nonumber &
2914208629378623635233568689620661059292824286678332244141710193543224&\\\nonumber &
6398913445666715611821217717113247490230785084573490330315258385027092&\\\nonumber &
8430635189422858159556640242387648505118727279976712594080826305686497&\\\nonumber &
5704192156828468122426913993820331040792083587374576453107288083533863&\\\nonumber &
7621042461928337773195702323394083899254945156232759245166073535525570&\\\nonumber &
0230213547360960426568078560686158288144399736900984300336175389023060&\\\nonumber &
5616437376142524810693077526916499349142045565890527378614647749168113&\\\nonumber &
6114052201940155433655070012448069659053921184482144683142113411245018&\\\nonumber &
2866313805439384828406861471681905442918834027976071244164504811135697&\\\nonumber &
3492188754652130466501198494962242554720665501900131522896323535118132&\\\nonumber &
6736727387287391416062134016119144495828759261687601071225293142451723&\\\nonumber &
2025574953555759702466842612215744635530149437609579838902237107080506&\\\nonumber &
2623093202608288457206286157404480893346053773771747450311895170916654&\\\nonumber &
6900489476391880896018139454661252438732214379026089121685763461541730&\\\nonumber &
5706473026557928820602015402938203420260179270235206050843670016447924.&
\end{align}

\end{document}